\documentstyle[twoside,fleqn,espcrc2,psfig]{article}
\newcommand{\bm}[1]{\mbox{\boldmath$#1$}}

\title{Perturbative renormalization parameters for heavy
  quarks\thanks{presented by K-I. Ishikawa}}
       
\author{
  K-I. Ishikawa\address{
    Department of Physics, Hiroshima University,
    Higashi-Hiroshima 739, Japan},
  S. Aoki\address{
    Institute of Physics,
    University of Tsukuba, Tsukuba, Ibaraki 305, Japan},
  S. Hashimoto\address{
    Computing Research Center, 
    High Energy Accelerator Research Organization(KEK),
    Tsukuba 305, Japan},
  H. Matsufuru$^{\rm a}$,
  T. Onogi$^{\rm a}$, and N. Yamada$^{\rm a}$
  }
       
\begin{document}

\begin{abstract}
  We study the heavy quark mass dependence of the
  perturbative renormalization factors for the heavy-light currents
  involving Wilson, Clover, and NRQCD heavy quarks. 
  We find that the one-loop $Z$-factor for the axial-vector
  current depends significantly on the heavy quark mass
  commonly for all actions we study, while that for the
  vector current has smaller dependence.
\end{abstract}

\maketitle

\section{Introduction}
Formulation of the heavy quark on the lattice, such as
NRQCD\cite{NRQCD} or the Fermilab approach with the
Wilson-like relativistic quarks\cite{FNAL}, requires
perturbative matching with the continuum theory. 
Since the quark mass in the lattice unit cannot be neglected 
for these theories, the perturbative calculation is much
more involved and have not developed as that for the
massless case.
We report the current status of our effort to calculate
the one-loop coefficients of the perturbative $Z$-factor
$Z_V$ and $Z_A$.
It is particularly important to calculate the mass
dependence of the axial-current $Z$-factor $Z_A$, which is
required for the calculation of the $B$ meson decay constant
$f_B$, since the one-loop coefficient in the static limit is
known to be very large\cite{static}.

The combinations of the heavy and light quark actions we
investigate in this report are listed in Table
\ref{tab:mmm}. 
The Wilson and Clover heavy quarks are used with the
Fermilab group's non-relativistic interpretation\cite{FNAL},
with which the relativistic actions can be treated in
parallel with NRQCD for heavy quarks, since the both methods 
are based on the small spatial momentum expansion, or
equivalently the $\Lambda_{QCD}/M$ expansion.
And in particular the both methods have the identical static 
limit.

The earlier works for the lattice perturbative calculation
for NRQCD can be found in Refs.\cite{NRPER,JSHEP}. 
Advanced studies using NRQCD-Clover actions
are also reported by J.~Shigemitsu and
A.~Ali-Khan\cite{JSAAH}.  

\begin{table}[tb]
  \caption{The combinations of heavy and light quark actions.}
  \label{tab:mmm}
  \begin{tabular}{cc}            \hline
    Heavy quark & Light quark  \\ \hline
    NRQCD       & Wilson       \\
    Wilson      & Wilson       \\
    Clover      & Clover       \\ \hline
  \end{tabular}
  \vspace{-1em}
\end{table}

\section{Operator Matching} 
Let us present the operator matching problem taking the
axial-vector current as an example.
The matrix element of the time component of the
axial-vector current $A_0$ for the free heavy and light
quark external fields having momenta $\bm{p}$ and $\bm{k}$
have the following form: 
\begin{eqnarray}
  \langle\bm{k}|J_{\mbox{\tiny N/W/C}}|\bm{p}\rangle & = &
  \bar{u_{l}}(k) \Bigl[ ( C_1^{(0)} + \alpha_s C_1^{(1)} )
  \gamma_5\gamma_4 
  \nonumber \\
  & + & a ( C_2^{(0)} + \alpha_s C_2^{(1)}) \gamma_5\gamma_4
  i\bm{\gamma}\cdot\bm{p} 
  \nonumber \\
  & + & a \alpha_s C_3^{(1)}
   i\bm{\gamma}\cdot\bm{k} \gamma_5\gamma_4  
  \nonumber \\
  & + & \mbox{higher orders}
  \left. \right]h
\end{eqnarray}
where $u_{l}$ is a light quark 4-component spinor, $h$ is a heavy
quark 2-component spinor, and the coefficients $C_1^{(i)},
C_2^{(i)} \dots $ are functions of the bare quark mass $a M_0$.
The first term gives the leading contribution surviving in
the static limit, and the second and third terms contribute
to the $\Lambda_{QCD}/m_{Q}$ and $a\Lambda_{QCD}$ corrections.

We have to match the continuum ($\overline{\mbox{MS}}$) and
the lattice theories to give the same matrix elements.
The matching of the tree level coefficients $C_1^{(0)}$ and
$C_2^{(0)}$ is almost trivial, and we have calculated in
this study the one-loop correction $C_1^{(1)}$, which gives
the multiplicative $Z$-factor $Z_A$, for all
combinations of the heavy and light quark actions.
The remaining coefficients $C_2^{(1)}$ and $C_3^{(1)}$ must
be calculated to achieve higher accuracy.

\section{NRQCD-Wilson}
We use two types of the NRQCD action including terms through 
$O(\Lambda_{QCD}/M)$ for the heavy quark.
The actions are given by
\begin{eqnarray}
S &= & \sum_{t,\bm{x}}Q(t,\bm{x})
\left[ Q(t,\bm{x}) - 
\left( 1-\frac{a H_{0}}{2 n} \right)^{n} \right. \nonumber \\
  && \times\left( 1-\frac{a \delta H}{2} \right)
           U^{\dagger}_{4}
           \left( 1-\frac{a \delta H}{2} \right) \nonumber \\
  && \left. \times \left( 1-\frac{a H_{0}}{2 n} \right)^{n}
  Q(t-1,\bm{x}) \right], \hspace{1mm}\mbox{type $A$},
\label{eqn:A}
\end{eqnarray}
and
\begin{eqnarray}
S &= & \sum_{t,\bm{x}}
\left[ Q(t+1,\bm{x})
\left( 1-\frac{a H_{0}}{2 n} \right)^{-n} U_{4}
\right. \nonumber\\
&& \times \left( 1-\frac{a H_{0}}{2 n} \right)^{-n}
   Q(t,\bm{x}) \nonumber \\
&& - Q(t,\bm{x})( 1 - a\delta H )Q(t,\bm{x}) \Bigr] ,
\hspace{2mm}\mbox{type $B$},
\label{eqn:B}
\end{eqnarray}
where $Q$ is the effective 2-component spinor field.
$H_{0}=-\Delta^{(2)}/[2 M_0]$ and 
$\delta H = -g\bm{\sigma}\cdot\bm{B}/[2M_0]$ are defined as
usual. 
The original 4-component spinor
field $\psi$ is related via Foldy-Wouthuysen-Tani
transformation through $O(\Lambda_{QCD}/M)$:
\begin{equation}
\psi = \left( 1 - \frac{\bm{\gamma}\cdot\bf{\Delta}}{2 M_0}\right)
\left(
\begin{array}{c}
Q \\
0      
\end{array}
\right)\equiv R h.
\end{equation}
Then the lattice current operators at tree level can be
written as
\begin{eqnarray}
J_{\mbox{\tiny N}} &=& \bar{q_{l}} \Gamma h - \frac{1}{2M_0}\bar{q_{l}}
      \Gamma(\bm{\gamma}\cdot\bm{\Delta}) h \nonumber \\
  &\equiv& J^{(0)}_{\mbox{\tiny N}}+J^{(1)}_{\mbox{\tiny N}}.
\end{eqnarray}

\begin{figure}[tb]
  \vspace{-3.5em}
  \begin{center}
    \leavevmode\psfig{figure=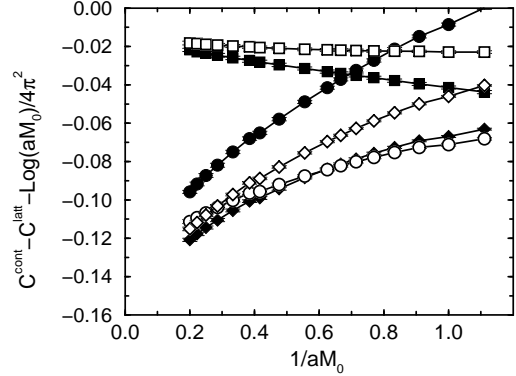,width=7.5cm,angle=-90}
  \end{center}
  \vspace{-3em}
  \caption{One-loop coefficients for the vector and
    axial-vector current renormalization factors for the
    type $A$ action.
    ${\scriptscriptstyle \bigcirc}$: time component of the
    axial vector current;
    $\Box$: time component of the vector current;
    $\Diamond$: spatial component of the vector current.
    Open symbols are obtained with the leading term
    $J_{\mbox{\tiny N}}^{(0)}$ only and filled symbols are
    with $J_{\mbox{\tiny N}}=J_{\mbox{\tiny N}}^{(0)} 
    +J_{\mbox{\tiny N}}^{(1)}$. }
  \label{Fig1}
  \vspace{-1em}
\end{figure}

We calculate the multiplicative part of current
renormalization factors
\begin{equation}
Z = 1 + g^{2}(C^{\mbox{\small cont}}-C^{\mbox{\small latt}}),
\end{equation}
where $C^{\mbox{\small cont}}$ and $C^{\mbox{\small latt}}$
are the one-loop corrections for the continuum and lattice 
(axial-)vector currents.
Figure \ref{Fig1} shows the $1/M_0$ dependence of
$C^{\mbox{\small cont}}-C^{\mbox{\small latt}}-\log(aM_0)/4\pi^2$.
We have applied the tadpole improvement using the average
plaquette for the mean link variable.
The axial-vector current in the continuum ($\overline{\mbox{MS}}$
scheme) is defined with the totally anti-commuting
$\gamma_5$. 

We observe that the time-component of the axial-vector
current and the spatial component of the vector current have 
a large $1/M_0$ dependence, and in the static limit we
reproduce the value $-$0.1346 for the axial-vector
current\cite{static}. 
This large dependence has an impact on the calculation of
$f_B$. 
For a typical value of the lattice spacing the heavy quark
mass becomes $aM_b$=1.8--2.5, and the difference of $Z_A$
from the static value could be large as $\sim$ 10\%.
The mass dependence of the time component of the vector
current is rather mild, on the other hand.
A similar behavior is observed for the type $B$ action.

\section{Wilson/Clover}
The renormalization factors for the Wilson-Wilson and
Clover-Clover heavy-light current are calculated in a
similar manner as the NRQCD-Wilson case.
The lattice current operator is written at tree level as
\begin{eqnarray}
 J_{\mbox{\tiny W/C}} &=& \bar{q_{l}}\Gamma q_{h} +
     d_1 \bar{q_{l}}\Gamma(\bm{\gamma}\cdot\bm{\Delta}) q_{h} \nonumber \\
   &=& J^{(0)}_{\mbox{\tiny W/C}}+ J^{(1)}_{\mbox{\tiny W/C}},
\label{eqn:currwc}
\end{eqnarray}
where $d_1=am_0/[2(2+am_0)(1+am_0)]$ with $am_0$ the bare
heavy quark mass appearing in the action.
In the non-relativistic interpretation approach the heavy
quark kinetic mass $am_2$ is given by $am_2=e^{am_1}\sinh
am_1/(1+\sinh am_1)$ and $am_1=\log(1+am_0)$.
The coefficient $d_1$ is small ($\sim$ 0.1 or smaller)
around the b-quark mass region and we neglect in this
study. 

\begin{figure}[tb]
  \begin{center}
  \vspace{-3.5em}
    \leavevmode\psfig{figure=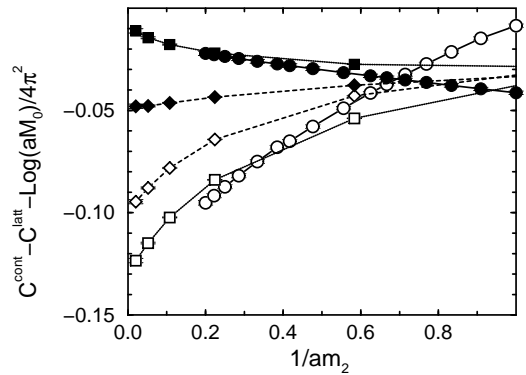,width=7.5cm,angle=-90}
  \vspace{-3em}
  \end{center}
  \caption{The heavy quark mass dependence of the renormalization
    factors for the time component of axial vector currents and
    vector currents. 
    ${\scriptscriptstyle \bigcirc}$ : NRQCD-Wilson;
    $\Box$ : Wilson-Wilson;
    $\Diamond$ : Clover-Clover.
    Open and filled symbols are for the axial vector
    current and the vector current respectively.
    }
  \label{Fig2n}
  \vspace{-1em}
\end{figure}

The one-loop coefficient for the time component of the
vector and axial-vector currents is shown in Figure
\ref{Fig2n} for Wilson-Wilson and Clover-Clover as well as
NRQCD-Wilson. 
We apply the tadpole improvement by using the critical
hopping parameter to define the mean link variable for the
relativistic actions.
Overall tendencies of the mass dependence are similar for
all actions, except in the lighter heavy quark mass
region, where the results for the relativistic actions tend
to have smaller mass dependence while the NRQCD-Wilson
result still has large slope.
The mass dependences for the vector current are weak compared
with that for the axial-vector current.

\section{Summary}
We have investigated the heavy quark mass dependence of
multiplicative part of the renormalization factors for the
heavy-light currents systematically using NRQCD, Wilson, and
Clover actions.
We found the large heavy quark mass dependence for $Z_A$

The calculation of the heavy-light decay constant is
presented in Ref.\cite{yamada} for NRQCD and in
Ref.\cite{hashimoto} for the relativistic actions.

\vspace{5mm}
This work is supported in part by Grants-in-Aid of the
Ministry of Education under the contract numbers 08044089
and 09740226.
H.M. is supported by the JSPS for Young Scientists for a
research fellowship.

\end{document}